\documentclass[aps,prl,twocolumn, groupedaddress]{revtex4-1}
\usepackage{graphicx}
\usepackage{amsmath}
\usepackage[T1]{fontenc}
\usepackage{mathtools}
\usepackage{siunitx}
\usepackage{lipsum}
\usepackage{amssymb}
\usepackage{xcolor}
\usepackage{lipsum}

\begin{document}


\title{ Prediction of Coupled Electronic and Phononic Ferroelectricity in Strained 2D h-NbN: First-principles Theoretical Analysis}


\author{Anuja Chanana and Umesh V. Waghmare}
\affiliation{Theoretical Sciences Unit, School of Advanced Materials and Sheikh 
Saqr Laboratory, Jawaharlal Nehru Centre for Advanced Scientific Research, Bangalore 560064, India}


\date{\today}

\begin{abstract}
Using first-principles density functional theoretical analysis, we 
predict coexisting ferroelectric and semi-metallic states in two-dimensional 
monolayer of h-NbN subjected to electric field and in-plane strain ($\epsilon$). 
At strains close to $\epsilon$=4.85\%, where its out-of-plane spontaneous polarization 
changes sign without inverting the structure, we demonstrate a hysteretic 
response of its structure and polarization to electric field, and uncover a 
three-state (P=$\pm$P$_o$, 0) switching during which h-NbN passes through Dirac 
semi-metallic states. With first-principles evidence for a combination of electronic and phononic ferroelectricity, we present a simple model that captures the energetics of coupled  
electronic and structural polarization, and show that electronic ferroelectricity 
arises in a material which is highly polarizable (small bandgap) and exhibits a 
large electron-phonon coupling leading to anomalous dynamical charges. 
These insights will guide search for electronic ferroelectrics, and our 
results on 2D h-NbN will stimulate development of piezo-field effect transistors 
and devices based on the multi-level logic. 
\end{abstract}


\maketitle

The symmetry of a crystalline material governs its functional properties. As a consequence of broken inversion symmetry, ferroelectric insulators exhibit spontaneous polarization that responds nonlinearly to external stress or electric fields. Ferroelectric polar order in perovskite oxide films gets suppressed or changes to inhomogeneously polarized states with diminishing film thickness, making ferroelectricity in ultrathin films 
fundamentally interesting \cite{junquera2003critical, fong2004ferroelectricity, 
gruverman2009tunneling, lee2015emergence, shukla2016anomalous}. Recent
development of a number of 2D materials has opened up new paradigm to
address fundamental size limits on ferroelectricity. For example,
improper ferroelectricity 
with out-of-plane polarization was predicted to emerge at a metal-semiconductor transition in 2D $1T-$MoS$_2$ \cite{shirodkar2014emergence}. On the other hand, ferroelectricity, with in-plane polarization has been shown to be stable in a number of 2D materials, like phosphorene \cite{wu2016intrinsic} and related Group V monolayers \cite{xiao2018elemental}, 
SnTe \cite{chang2016discovery} and other Group-IV monochalcogenides \cite{PhysRevLett.117.097601, wan2017promising}, $\beta$-GeSe \cite{PhysRevB.97.144104}, SbN and BiP \cite{liu2018robust} and GaTeCl \cite{zhang2018controllable}. Recently, out-of-plane ferroelectric order was predicted in AgBiP$_2$Se$_6$ \cite{xu2017monolayer} and In$_2$Se$_3$ \cite{ding2017prediction}. Ferroelectricity in these 2D materials arises mostly from phonons, except in $1T-$MoS$_2$, where strongly coupled electrons and phonons are involved \cite{shirodkar2014emergence}. 

Electronic ferroelectricity in 3D crystals is relatively 
rare \cite{kobayashi2012electronic}, and  yet to be explored in 
2D materials. Many 2D materials exhibit a small 
electronic band-gap and a variety of centrosymmetric and non-centrosymmetric structures.
Recently, 2D polymorphs of monolayered Niobium Nitride, s-NbN and h-NbN 
were predicted theoretically based on first-principles calculations \cite{NbN_1}. 
While s-NbN is metallic and retains the superconducting properties of bulk NbN 
at low T, h-NbN is a piezoelectric semiconductor manifesting a rare sd$^2$-sp$^2$ hybridization with an electronic structure that is similar to that of monolayered MoS$_2$. Structure of
h-NbN is non-centrosymmetric, and it exhibits an out-of-plane polarization
due to buckled structure (Nb and N atoms are in distinct planes). However, its polarization is not readily switchable with electric field as the buckling is too large. Piezoelectric coupling of h-NbN \cite{NbN_1} means that its polar buckling can be reduced with strain, and ferroelectricity could be practically achievable with in-plane strain. Secondly, the unusual hybridization in its electronic structure makes h-NbN a promising material to host electronic ferroelectricity.

In this letter, we present the exploration of ferroelectricity in monolayered h-NbN subjected to in-plane lattice strain ($\epsilon$$_{xx}$=$\epsilon$$_{yy}$=$\epsilon$) using first-principles density functional theoretical analysis of electronic, structural and vibrational properties,  polarization and associated potential discontinuity. We show that strained h-NbN is locally stable upto $\epsilon$=5.3\%, beyond which it yields  to a cell doubling instability, and transforms to the planar structure with vanishing buckling ($\Delta$=$d^{\mathrm{z}}_{\mathrm{Nb}}$-$d^{\mathrm{z}}_{\mathrm{N}}$) for $\epsilon$$\geqslant$6.5\%. Spontaneous polarization of h-NbN changes its sign at $\epsilon$=4.85\% without undergoing structural inversion (a change in the sign of $\Delta$). Thus, the window of strain $\epsilon$ $\in$ (4.85-5.3)\% is rather interesting for ferroelectricity originating from electrons and phonons. We demonstrate electric field driven polarization switching in h-NbN at  $\epsilon$=5\%, and present a simple model to capture the physics of its strongly coupled electronic and phononic ferroelectricity.

Our first-principles calculations are based on Density Functional Theory (DFT) within a generalized gradient approximation (GGA) \cite{PhysRevLett.77.3865} of the exchange-correlation energy with a functional parametrized by Perdew-Wang \cite{perdew1986accurate}, as implemented in Quantum Espresso package \cite{QE-2009}. Vanderbilt ultrasoft scalar relativistic pseudopotentials have been used to describe the interaction  between ionic cores and valence electrons \cite{PhysRevB.41.7892}. We represent the electronic wave functions with a plane wave basis truncated at an energy cut-off of 40 Ry. Similarly, charge density is represented in the basis set truncated at energy cut-off of 400 Ry. A uniform mesh of 20x20x1 k-points was used for sampling Brillouin zone integrations. We relax each structure using Hellman-Feynman forces and Broyden Fletcher Goldfarb Shenno (BFGS) scheme till the force on each atom $\overrightarrow{F}$ is less than 10$^{-3}$ Ry/Bohr. Fermi-Dirac distribution with a width of k$_B$T=0.005 Ry is used for smearing occupation of numbers of electronic states. We use a large vacuum (thickness of 50\AA) along the z-direction to ensure weak interactions between the periodic images of the monolayer. Dynamical matrices are calculated using Density Functional Perturbation Theory (DFPT) at wave vectors (q) on a mesh of 6x6x1 in the BZ, and Fourier interpolated to obtain phonon dispersion. We used finer mesh of q points in estimation of the electron-phonon couplings.  We determine spontaneous polarization using Berry phase method \cite{king1993theory}. Using a vacuum thickness of 50\AA \hspace{0.07 cm} and including dipole corrections to eliminate the effects of polar field arising from the continuity and periodicity of electrostatic potential, we check the validity of our results. To simulate the response of h-NbN to electric field, we add a saw-tooth potential as a function of z, and simulated effects of electric field varying from -0.6 to 0.6V/\AA \hspace{0.07 cm} in the intervals of 0.1V/\AA, and from -6 to 6V/\AA \hspace{0.07 cm} in the intervals of 1V/\AA \hspace{0.07 cm}. To simulate hysteresis, we monotonously increase or decrease electric field in these steps, and use the relaxed structure from the previous step as the initial structure during structural relaxation at each field. 

From the energies of buckled and planar ($\Delta$=0\AA \hspace{0.07 cm}) structures of h-NbN (Fig. 1(a)) as a function of $\epsilon$, we find that the energy of the planar structure decreases with tensile $\epsilon$, and has a minimum energy at $\epsilon$=8\%. On the other hand, the energy of buckled structure increases upto $\epsilon$=6.5\% and drops sharply beyond that, transforming spontaneously to the planar structure (see Fig. S1: where the planar structure has an energy lower than the buckled structure at $\epsilon$$\textgreater$5\%; while there is an energy barrier that separates the two upto $\epsilon$$\doteq$6.5\%). The two energy curves of buckled and planar structures intersect at $\epsilon$=5.26\% corresponding to a lattice constant of 3.32\AA \hspace{0.07 cm} marking a transition from  buckled to the planar structure. To study the structural transition at $\epsilon$>6.5\% from buckled to planar structure, we analyze their electronic structures (see Fig. S2) at $\epsilon$=6.5\% and $\epsilon$=7\% by interchanging their lattice constants, but maintaining the buckling, and find that the structural transition is due to change in the buckling of the h-NbN structure. The electronic structures of the original structure and the structure with interchanged lattice constants are shown in Fig. S2. 

To analyze the dynamical stability of strained h-NbN, we determine its phonon spectrum. From $\omega$ of the zone center optical mode and M-point acoustic modes of h-NbN as a function of $\epsilon$ (see inset of Fig. 1(a)), we see that frequencies reduce with tensile strain $\epsilon$ and drop sharply for $\epsilon$$>$5\%, M-point mode becoming unstable at $\epsilon$$\geq$5.3\% (see Fig. S3). In the phonon dispersion of h-NbN at $\epsilon$ from 5\%-6.5\% (Fig. S3), we note that the lowest frequency branches of acoustic and optical modes shift downwards with $\epsilon$, with frequency at M-point becoming negative for $\epsilon$>5.3\%. We find that the spontaneous polarization of buckled h-NbN depends strongly on $\epsilon$ (see Fig. 1(b)): it increases upto $\epsilon$=6.5\% and drops sharply to 0 pC/m \hspace{0.07 cm} from $\epsilon$=7\%-8\%, confirming its transformation from buckled to planar structure. Most notably, polarization changes its sign (indicated by a blue dashed line in Fig. 1(d)) at $\epsilon$=4.85\%, and a corresponding change is seen in the macroscopic average potential (see Fig. S4), while  $\Delta$ retains its sign, i.e. P switches its sign without inversion of h-NbN structure. This is the first hint to possible electronic ferroelectricity in h-NbN. Secondly, a rather small value of P (0.248 pC/m in comparison with the polarization of other 2D materials \cite{wu2016intrinsic, xiao2018elemental, chang2016discovery, PhysRevLett.117.097601, wan2017promising, PhysRevB.97.144104, liu2018robust,zhang2018controllable,xu2017monolayer,ding2017prediction}) inspite of significant structural polarity evident in large buckling $\Delta$ (-0.637\AA) suggests that electronic polarization may be strongly cancelling the ionic polarization. In the strain window $\epsilon$ $\in$ [4.7\%-5.31\%], where the spontaneous polarization switches direction and the structural stability is maintained, we choose h-NbN strained at $\epsilon$=5\% for exploration of its ferroelectricity. The strongly buckled structure of h-NbN makes it capable of sustaining large in-plane strain. 

\begin{figure}[htpb]
\includegraphics[width=9cm,height=12cm,keepaspectratio]{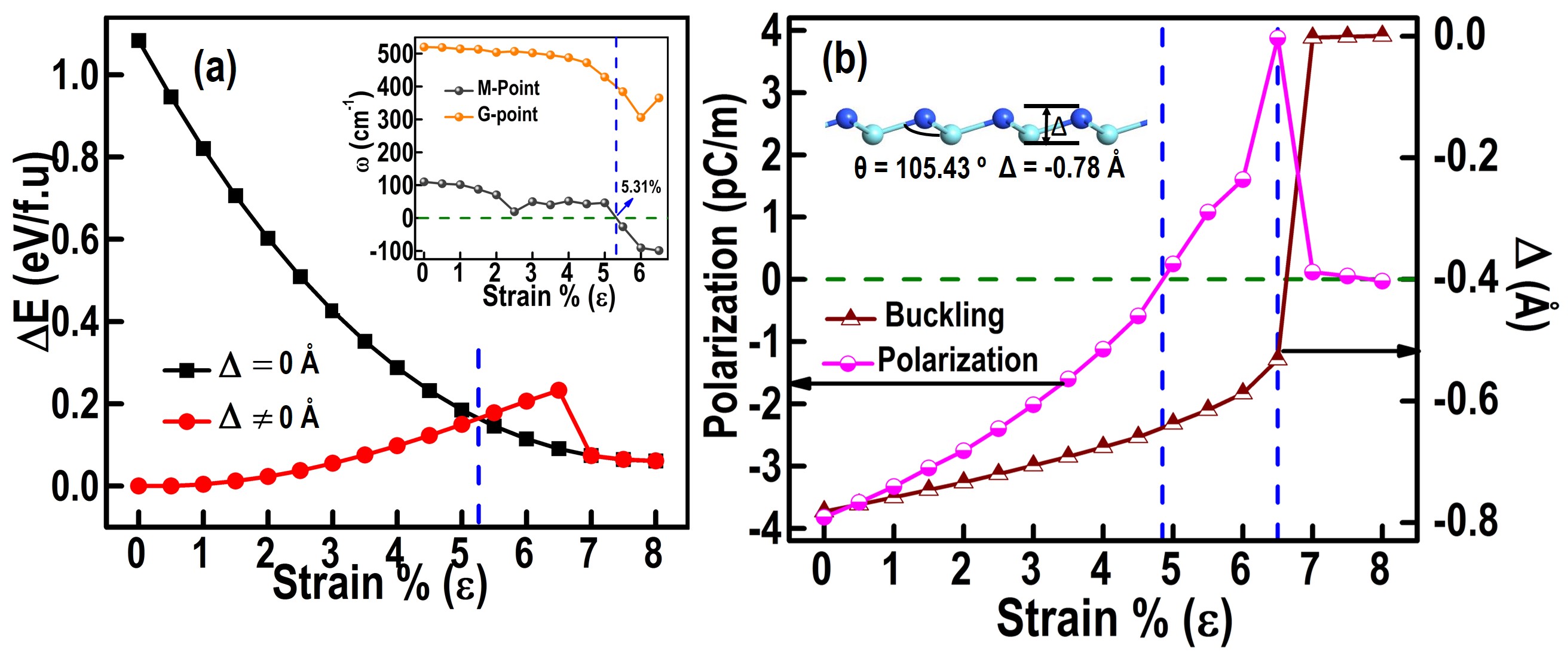}
\caption{(a) Energies of planar and buckled structures of h-NbN as a function of $\epsilon$, which cross at $\epsilon$=5.26\%. Inset shows the dependence of frequencies of $\Gamma$ optical and M acoustic modes as a function of $\epsilon$, showing a structural instability emerging at $\epsilon$=5.31\%. (b) Electric polarization and buckling ($\Delta$=$d^{\mathrm{z}}_{\mathrm{Nb}}$ - $d^{\mathrm{z}}_{\mathrm{N}}$) as a function of $\epsilon$. Inset shows the structure of monolayered h-NbN. The sign of electric polarization reverses at $\epsilon$=4.85\%, while the sign of buckling $\Delta$ is unchanged. $\Delta$ vanishes for $\epsilon$$\geqslant$6.5\%, and P vanishes too.}
\end{figure}

We begin by examining the electronic and vibrational structure of h-NbN at $\epsilon$=5\%.  It has a direct band gap of 0.16eV along the K $\longrightarrow$ $\Gamma$ path (see Fig. 2(a)), noticeably lower than that of unstrained h-NbN (0.77eV) \cite{NbN_1}. Its phonon dispersion (Fig. 2(b)) exhibits no unstable modes, establishing its dynamical stability. Its optical and acoustic phonon bands are well separated by a gap of 262 cm$^{-1}$, a bit narrower than that of unstrained h-NbN (300 cm$^{-1}$) \cite{NbN_1}. Electronic and vibrational structure of h-NbN at $\epsilon$=4.7\% (see Fig. S7) are similar, with a direct electronic bandgap of 0.2 eV and a vibrational gap of 267 cm$^{-1}$. Energy of h-NbN at $\epsilon$=5\% as a function of buckling has a shape of a triple-well (see Fig. 2(c)). The two valleys denoted by 1 and 5 correspond to energy minima of its ferroelectric states with opposite P. The transition states (2 and 4) are unstable, and give the energy barrier to be crossed during homogeneous switching of P. The 3$^{rd}$ valley at $\Delta$=0 \AA \hspace{0.07 cm} corresponds to its planar structure. Buckling and energetics of triple-well potential of h-NbN at $\epsilon$=4.7\% and $\epsilon$=5\% are similar (see Table S1), and comparable to the energy barriers of other 2D materials. While the buckling $\Delta$ is quite large ($\backsim$ -0.63 and -0.65 \AA \hspace{0.07 cm}) in h-NbN (at both $\epsilon$=5\% and 4.7\%), its polarization see Fig. 2(d) (i) is quite small, and (ii) has a sign that is switchable with both strain ($\epsilon$) and buckling ($\Delta$). Since the energy profile of h-NbN describes a triple-well potential, we expect an unconventional switching of its structure. 

\begin{figure}[htpb]
\centering
\includegraphics[width=9cm,height=12cm,keepaspectratio]{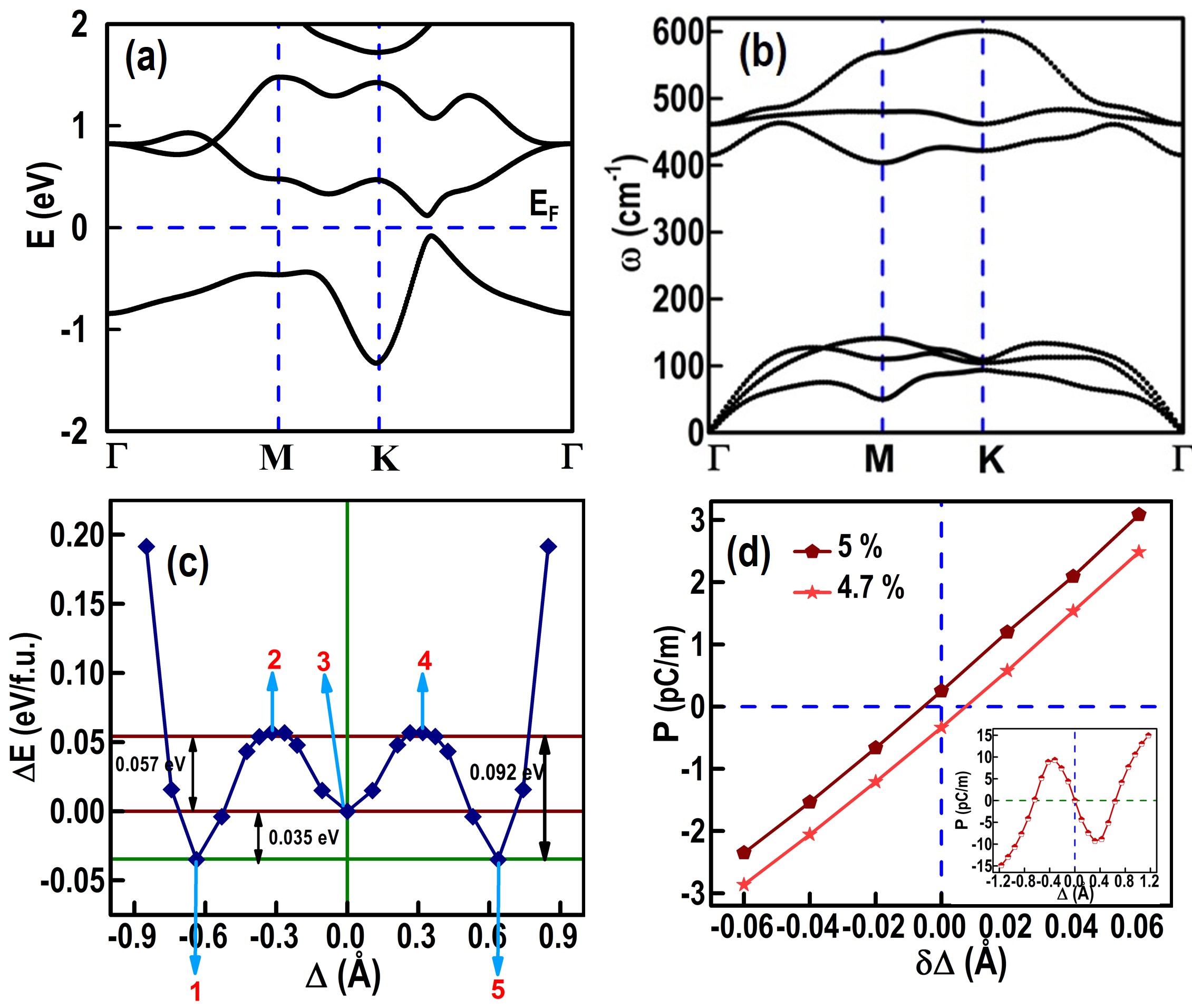}
\caption{(a) Electronic structure, (b) phonon dispersion and (c) energetics of buckling ($\Delta$) that is relevant to polarization switching in h-NbN at $\epsilon$=5\%. In (c), energy of the planar structure of h-NbN (configuration 3) is the reference. Ferroelectric states correspond to configurations 1 and 5, and the states 2 and 4 are the transition states giving energy barriers to be crossed during switching. (d) Polarization as a function of change in buckling of h-NbN (configuration 5 in (c)) at $\epsilon$=5\% and $\epsilon$=4.7\%, revealing that the sign of P reverses without inversion of the structure {\it i.e.} change in sgn($\Delta$). Inset shows that  polarization changes sign with $\Delta$, and vanishes for the flat structure.  }
\end{figure}

\begin{figure*}[htpb]
\centering
\includegraphics[width=16cm,height=12cm,keepaspectratio]{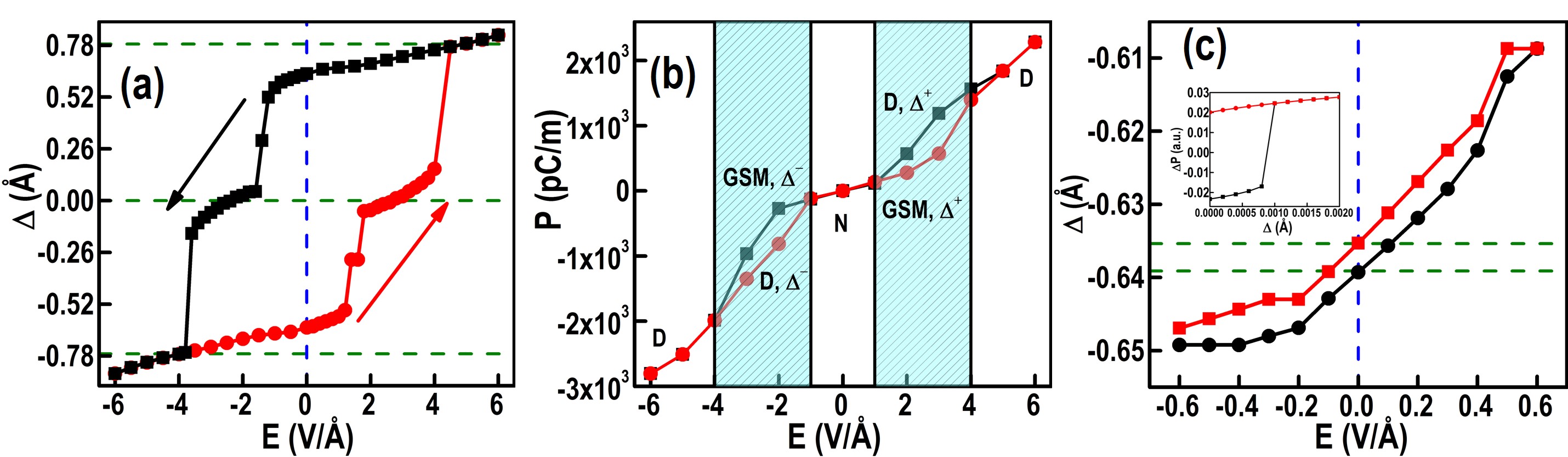}
\caption{Response of h-NbN ($\epsilon$=5\%) to electric field (a) hysteretic switching of buckling ($\Delta$=$d^{\mathrm{z}}_{\mathrm{Nb}}$-$d^{\mathrm{z}}_{\mathrm{N}}$) as a function of electric field (high values), (b) corresponding polarization exhibiting hysteretic behavior. Here, $D, \Delta^+$ ($D, \Delta^-$) label a Dirac semi-metallic state with a positive (negative) $\Delta$, and GSM indicates a gapped semi-metallic state. $D$ labels Dirac semi-metallic state and N is the equilibrium semiconductor. A weak hysteresis in buckling response to low electric field (c), and polarization as a function of buckling obtained using our model Hamiltonian (inset of Fig. 3(c)). The arrows in (a) indicate the directions of changes in electric field along two paths, and the same symbols are used in other panels.}
\end{figure*}

We simulated the response of h-NbN at $\epsilon$=5\% to electric field (E) applied perpendicular to its plane. We start with E=-6V/\AA \hspace{0.07 cm} and relax the  structure with $\Delta$ ($d^{\mathrm{z}}_{\mathrm{Nb}}$-$d^{\mathrm{z}}_{\mathrm{N}}$)=-0.637\AA, and gradually increase E in steps of 1V/\AA. We find a nonlinearity in $\Delta$(E), with a two-step switching (see Fig. 3(a)). Buckling (red curve) increases in magnitude as E increases (red arrow), with first switch from $\Delta$=-0.55\AA \hspace{0.05 cm} to -0.053\AA \hspace{0.05 cm} at E $\sim$ 1.2V/\AA \hspace{0.07 cm} to 1.8V/\AA, followed by the second switch from $\Delta$=0.16 \AA \hspace{0.07 cm} to 0.76\AA \hspace{0.07 cm} at E $\sim$ 4V/\AA \hspace{0.07 cm} to 4.2V/\AA. Along the reversed path (black arrow) starting with E=6V/\AA, we see that the magnitude of buckling increases (black curve) linearly first. Here as well, we find a two-step switching in $\Delta$(E): first from $\Delta$=0.52\AA \hspace{0.07 cm} to 0.046\AA \hspace{0.07 cm} at E $\sim$ -1.2V/\AA \hspace{0.07 cm} to -1.6V/\AA \hspace{0.07 cm} and the second switch from $\Delta$=-0.166\AA \hspace{0.07 cm} to -0.76\AA \hspace{0.07 cm} at E $\sim$ -3.6V/\AA \hspace{0.07 cm} to -3.8V/\AA. At E=0V/\AA, \hspace{0.07 cm} $\Delta$=-0.637\AA \hspace{0.07 cm} (red curve) and 0.637\AA \hspace{0.07 cm} (black curve), correspond to the equilibrium buckling at  $\epsilon$=5\% (shown by vertical blue dashed lines). The structure at E $\in$ (1.2,4.5)V/\AA \hspace{0.07 cm} and E $\in$ (-3.8,-1.2)V/\AA \hspace{0.07 cm} corresponds to a nearly flat structure changing in response to electric field (vertical blue shaded regions).

We now discuss the correspondence between (Fig. 3(a) and Fig. 3(b)) polarization and buckling $\Delta$ as a function of electric field. For electric fields E<-4V/\AA \hspace{0.07 cm} and 5V/\AA \hspace{0.07 cm}<E, polarization obtained along the two paths (red and black curves) coincide, similar to the corresponding buckling at these electric fields. For 1<|E|<4V/\AA, we find a remarkable difference in P and in the electronic structure obtained along the two paths (two branches of the hysteresis curve). Along the forward path (red curve), it corresponds to $\Delta$=0\AA \hspace{0.07 cm}, (the flat structure), with electronic structure consisting of non-crossing valence and conduction bands that overlap in energy (gapped semi-metallic state). Along the reversed path, $\Delta$ $\backsim$ 0.52\AA,\hspace{0.07 cm} and its electronic structure is that of a Dirac semimetal (with Dirac like bands crossing from $\Gamma$ $\longrightarrow$ M and K $\longrightarrow$ $\Gamma$) (see Fig. S8). Thus, the two states of h-NbN at |E| $\in$ (1,4)V/\AA \hspace{0.07 cm} correspond to distinct electronic states (and buckling $\Delta$), revealing that switching occurs between inverted states that have qualitatively different electronic structure. In the SI (Fig. S9), we show that the spin-orbit coupling results in spin-split bands, transforming the DSM into triple point semi-metallic state. Close examination of the electronic structure of such DSM/TSM shows that Dirac point involves spin split bands of h-NbN, and a parabolic band of free carriers in the "electrode" modeled with saw tooth potential. Thus we term this as an "extrinsic" DSM/TSM state of h-NbN as  it depends on the electrode as well. 

While the ferroelectric state of h-NbN is gapped, it passes through semi-metallic states while switching its polarization (see Fig. 3(b)), and this is possible because its sub-angstrom thickness is smaller than its screening length in metallic states, which permitting electric field induced switching, similar to metallic ferroelectrics \cite{filippetti2016prediction, fei2018ferroelectric}. Thus, ferroelectricity in h-NbN has strongly coupled electronic and phononic contributions. At E $\in$ (-1V/\AA \hspace{0.07 cm}, 1V/\AA \hspace{0.07 cm}), polarization is rather small (see inset of Fig. 3(c)), with  $\Delta$ in this range of electric field being $\pm$0.637\AA \hspace{0.07 cm}, which is the buckling of h-NbN at $\epsilon$=5\%. Moreover, the electronic structure for E $\in$ (-1V/\AA \hspace{0.07 cm}, 1V/\AA \hspace{0.07 cm}) is similar to that of h-NbN at $\epsilon$=5\%. Such small magnitude of P is because the electronic and ionic contributions to P cancel each other rather effectively, and minimize the energy cost associated with depolarization field. We have resolved the ambiguity of polarization quantum, by ascertaining the P($\Delta$=0)=0, using Berry phase technique as well as from the discontinuity in macroscopically averaged potentials (see Fig. S10). We note that electric fields used in our simulation are significantly higher (because they induce homogeneous switching at T=0K) than that are achievable in realistic devices. We expect heterogeneous nucleation of domains to be active and consequent switching to be possible at much lower electric fields in experimental verification of ferroelectricity in h-NbN.

To understand this, we simulated the response of h-NbN at $\epsilon$=5\% to smaller values of E (-0.6V/\AA \hspace{0.07 cm} to 0.6V/\AA \hspace{0.07 cm}), and indeed see a weakly hysteretic response of buckling (see Fig. 3(c)) as a function of electric field. At 0V/\AA, $\Delta$=-0.639\AA \hspace{0.07 cm} (red curve) and -0.635\AA \hspace{0.07 cm} (black curve), which bracket the buckling $\Delta$=-0.637\AA \hspace{0.07 cm} at E=0. As the structure of h-NbN, remains in the close neighborhood of equilibrium structure ($\Delta$=-0.637\AA \hspace{0.07 cm}), this weak hysteresis is related to the switchable polarization shown in Fig. 2(d), and is driven primarily by electronic mechanism of ferroelectricity. Origin of electronic ferroelectric polarization can be traced to opposite changes in Lowdin charges of Nb and N in response to small electric fields (see Fig. S12(b)). We carried out a similar analysis of field-dependent hysteretic response of h-NbN at $\epsilon$=4.7\%  to establish the robustness of our conclusions of electronic and phononic ferroelectricity in h-NbN (see Fig. S13 and S14).

We now present a simple model that captures the physics of coupled
ferroelectric orders originating from electrons (P\textsubscript{e})
and from phonons (P$_i$=Z.u, u=$\Delta$-$\Delta$$_0$).
Our model expresses energy per unit area of h-NbN ($\epsilon$=5\%) as a Taylor series expansion
in these order parameters, with the ferroelectric ground state of h-NbN
(with a non-zero buckling, $\Delta$$_0$) as the
reference structure.
\begin{equation}
H(P_{e},u,E)=\frac{1}{2}A P_{e}^{2}+\frac{1}{4}CP_{e}^{4}-EP_{e}+\frac{1}{2}Ku^{2}-ZuE+\mu uP_{e},
\end{equation}
where, A(>0) is the stiffness associated with electronic
dipole, which is inversely proportional to electronic contribution
to dielectric susceptibility (A$\propto$$\frac{1}{\chi}$), C is a quartic coefficient, K is the spring constant associated with change in buckling u=$\Delta$-$\Delta$$_0$, Z is the Born charge, and $\mu$ is the dipolar electron-phonon coupling (EPC). We point out that the instability of electronic polarization is qualitatively unaffacted by the symmetry allowed terms in H (see Fig. S15). We first demonstrate the effect of electron-phonon coupling on
the descriptors of ferroelectricity {\cite{kobayashi2012electronic, waghmare2003first}}. Minimizing the energy with respect to electronic ferroelectric order, P$_e$, we get: 
\begin{equation}
K\rightarrow K-\frac{\mu^{2}}{A},Z\rightarrow Z-\frac{\mu}{\chi^{'}},
\end{equation}
where it is evident that the EPC leads to softening of phonon (reduction in K),
and renormalization of the Born charge to anomalous values (deviating
from the nominal charge). The former is directly relevant to the phonon-driven
ferroelectricity, while the latter is known to be an indicator of
electronic {\cite{kobayashi2012electronic}} as well as ionic 
ferroelectricity \cite{waghmare2003first}. A similar analysis can be carried out by minimization of energy with respect to u, noting that phonon (u) is stable (K$>$0).
If the electronic dielectric susceptibility is large (i.e., A
is small,  +ve), effects of nontrivial
electron-phonon coupling are significant It can be shown from the model, that electronic
ferroelectricity arises (renormalized A<0), if ($\mu$$^{2}$/{K}$)>$A 
i.e. if the electron-phonon coupling $\mu$ is large and A
is small. h-NbN, being a good superconductor, its electron-phonon coupling
is known to be large {\cite{NbN_1}}. Secondly, A
is small when a material is close to a semiconductor to metal phase transition
(vanishing bandgap), which is known here to occur in h-NbN at strain $\epsilon$=6\%.
In Fig. 3(b), the hysteretic loops are indeed associated with dominantly
electronic ferroelectricity (where the gap is close to zero in the
nearly planar structure, see Fig. 3(a)). At smaller electric fields,
the electronic hysteresis is weaker Fig. 3(c) (as A is larger
in the buckled structure as seen in Fig. 3(c)). Analysis
of our model reproduces the hysteresis associated with electronic ferroelectricity (see inset in Fig. 3c) at small fields.

We have established unusual ferroelectricity in strained 2D monolayer of h-NbN that originates from its electrons as well as phonons, with out-of-plane spontaneous polarization that couples with strain and electric field. At $\epsilon$=5\%, it exhibits a two-step structural switching through Dirac and other semi-metallic states that is controlled by external electric field, and governed by a triple-well energy landscape. We presented a simple model Hamiltonian based on Landau theory to study the coupled electronic and phononic ferroelectricity in h-NbN, and uncover the descriptors of electronic ferroelectricity which is rare in nature: \textit{high electronic polarizability and anomalous Born charges.} These descriptors will guide search for novel electronic ferroelectric materials. Our work opens up h-NbN as a realistic model system that can be explored to understand the conflicting behavior of a semiconductor to turn into metallic or polar ferroelectric states. 

\section{Acknowledgement}
UVW acknowledges support from a JC Bose National Fellowship of DST, government of India, Sheikh Saqr fellowship and IKST-KIST, Bangalore.

\bibliography{NbN_Strain}

\end{document}



\title{Prediction of Coupled Electronic and Phononic Ferroelectricity in Strained 2D h-NbN: First-principles Theoretical Analysis  Supplementary Information}


\author{Anuja Chanana and Umesh Waghmare}
\affiliation{Theoretical Sciences Unit, Jawaharlal Nehru Centre for Advanced Scientific Research, Bangalore 560064, India}

\date{\today}
\index{}
\printindex



\pacs{}

\maketitle


\section{}

%

\begin{figure}
\centering
\renewcommand{\thefigure}{S\arabic{figure}}
\includegraphics[width=9cm,height=12cm,keepaspectratio]{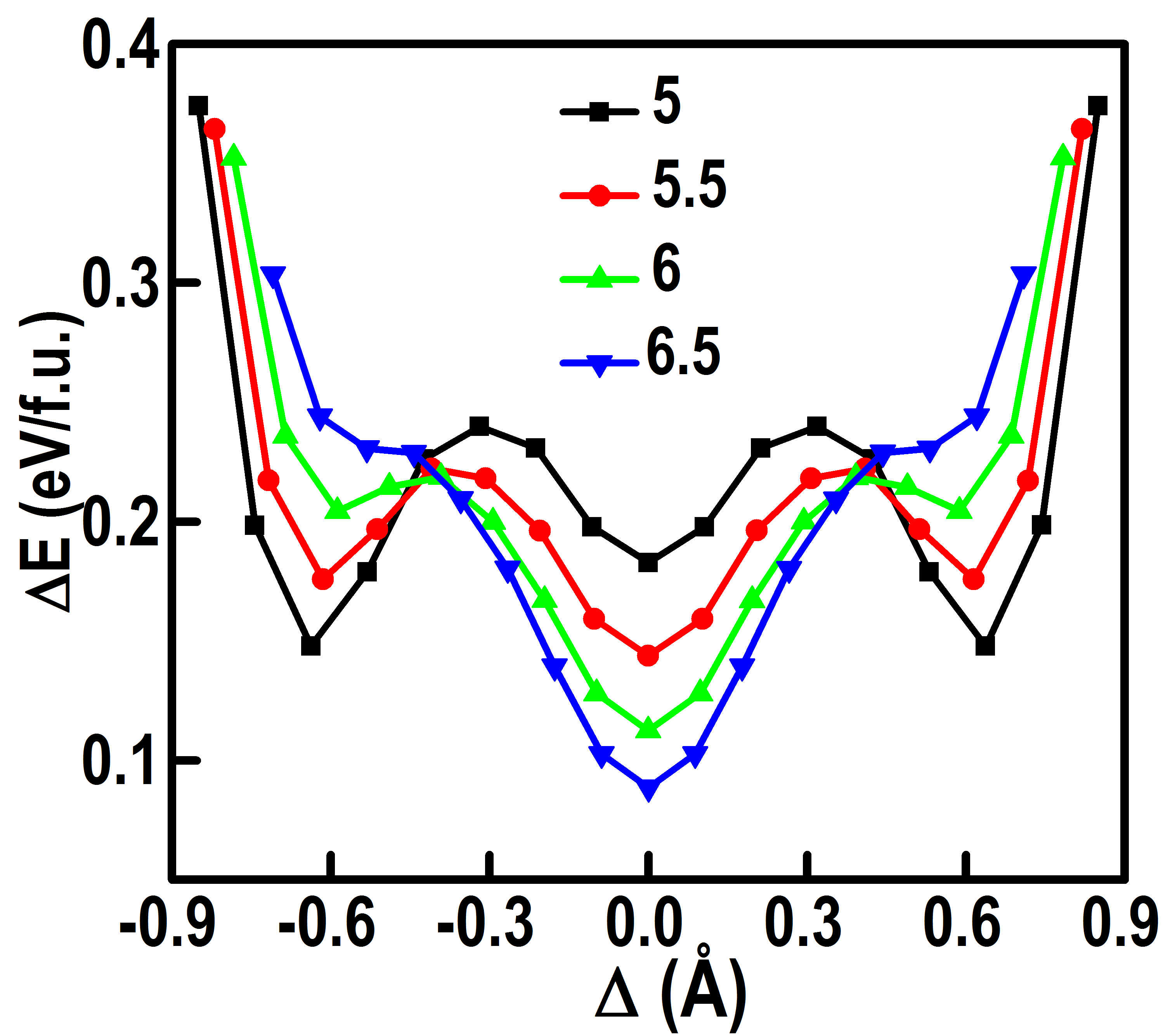}
\caption{Energetics of buckling of h-NbN for $\epsilon$ $\geq$ 5\%.}
\end{figure}

Fig. S1 shows energy as a function of buckling at $\epsilon$ $\geq$ 5\%. For $\epsilon$ $\leq$ 6.5\%, h-NbN in buckled configuration is meta-stable, and exhibits energetics in the form of triple-well potential. The planar configuration of h-NbN has energy minima for $\epsilon$ $\geq$ 5.5\%. \\

\begin{figure}
\centering
\renewcommand{\thefigure}{S\arabic{figure}}
\includegraphics[width=15cm,height=24cm,keepaspectratio]{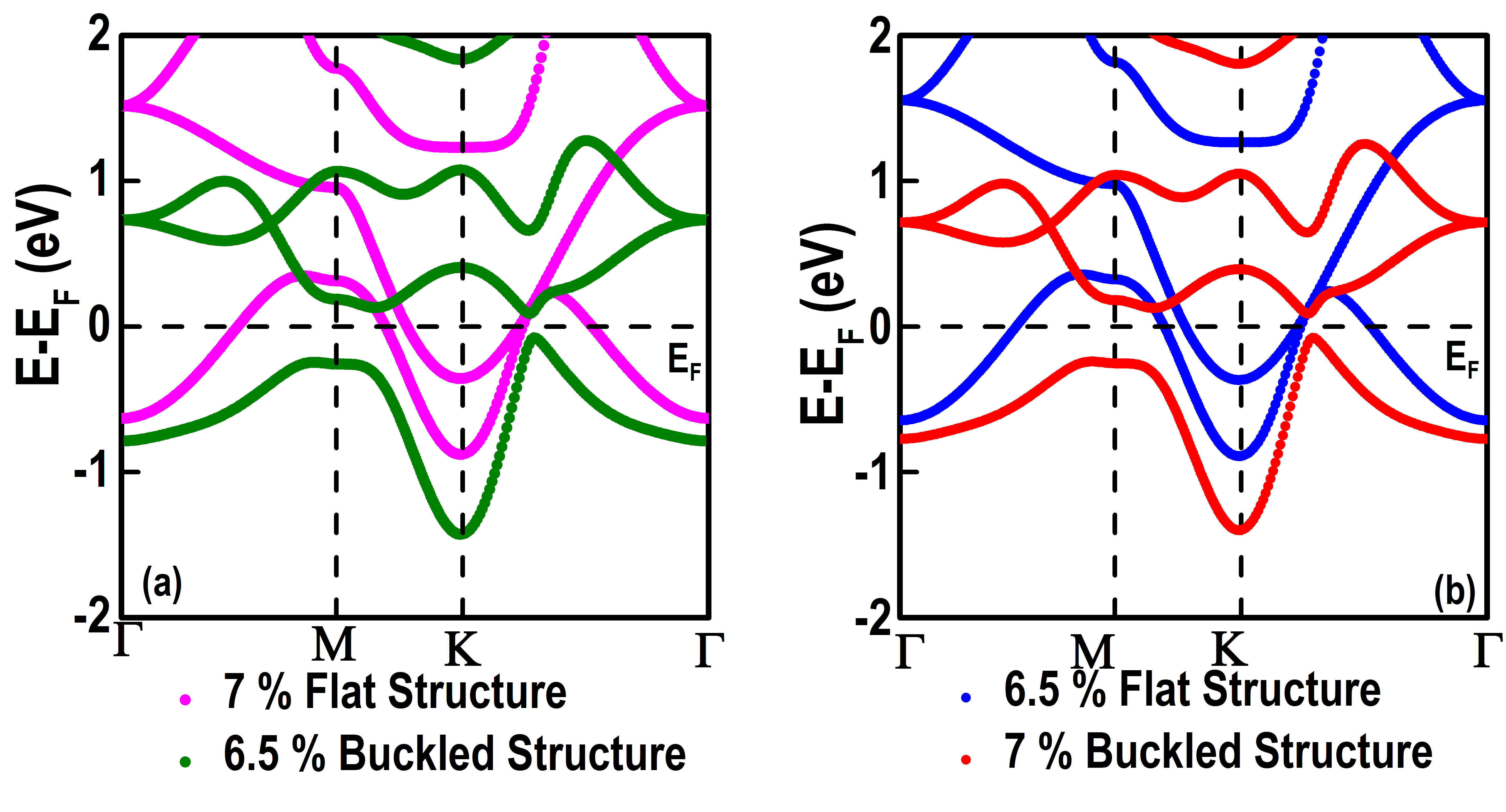}
\caption{Comparison of electronic structures of h-NbN by interchanging lattice constants at $\epsilon$ = 6.5\% (buckled structure) and 7\% (flat structure). }
\end{figure}

Fig. S2 compares the electronic structure of buckled ($\epsilon$ = 6.5\%) and flat ($\epsilon$ = 7\%) h-NbN structure. We interchange the lattice constants of structures at each $\epsilon$, maintaining their atomic positions and observe that the flat structure at $\epsilon$ = 7\% (pink curve) and $\epsilon$ = 6.5\% (blue curve) in Fig. S2 (a) and Fig. S2 (b) are identical to each other. We also observe similar nature for the buckled structure at $\epsilon$ = 6.5\% (green curve) (Fig. S2 (a)) and $\epsilon$ = 7\% (red curve) (Fig. S2 (b)). Thus, we establish that the semiconductor to metallic transition is caused by buckling in h-NbN structure from $\epsilon$ = 6.5\% - 7\%.\\

\begin{figure}
\centering
\renewcommand{\thefigure}{S\arabic{figure}}
\includegraphics[width=9cm,height=12cm,keepaspectratio]{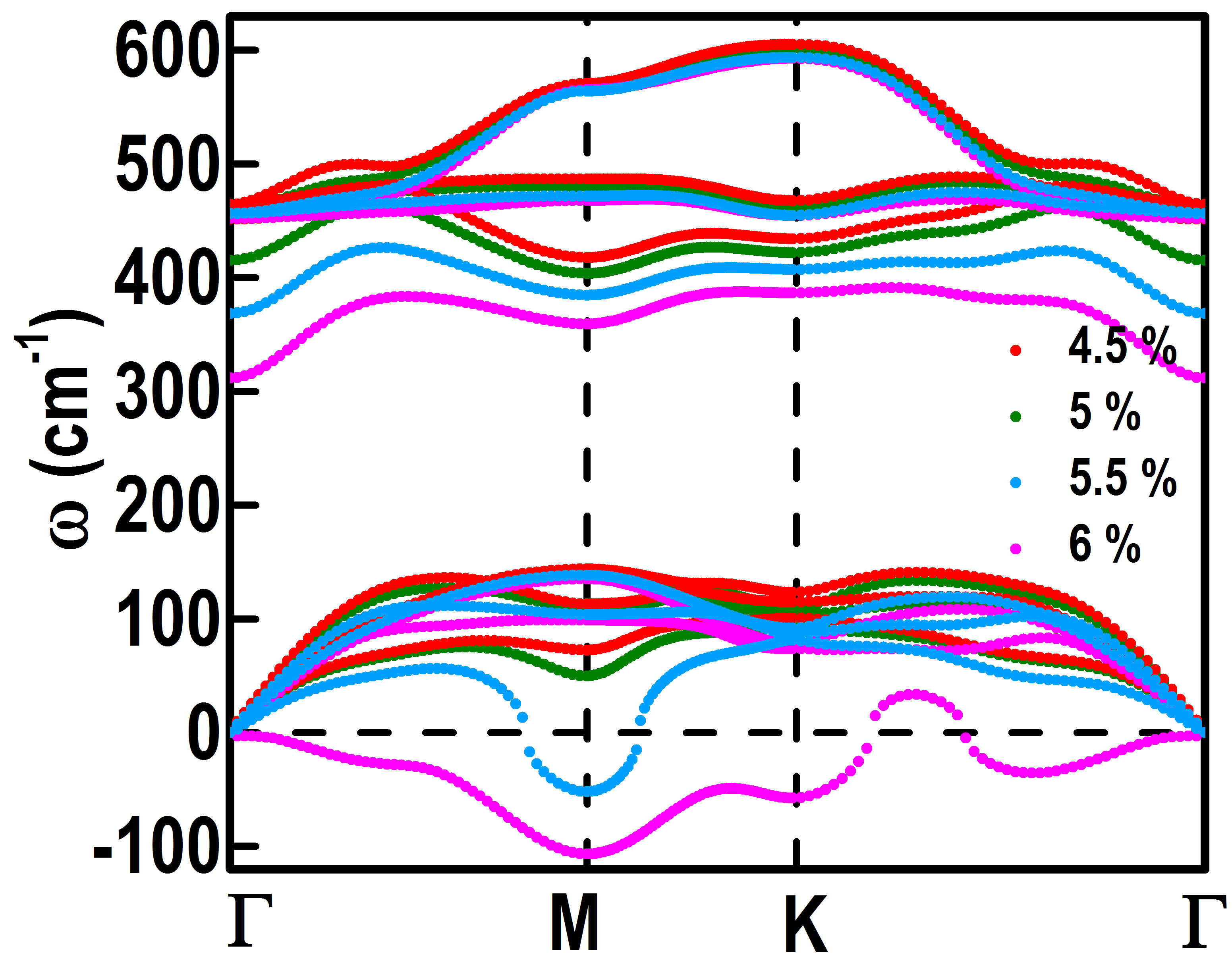}
\caption{Phonon dispersion of h-NbN with $\epsilon$ ranging from 4.5\% - 6\%.}
\end{figure}

Fig. S3 shows the phonon dispersion of h-NbN with $\epsilon$ ranging from 4.5\% - 6\% in the intervals of 0.5\%.  With an increase in $\epsilon$, the branches of phonon dispersion shift downwards and it is more significant for the lower branches of both acoustic and optical modes.  Moreover, the difference between the acoustic and optical modes also decreases from 266 cm$^{-1}$ at $\epsilon$ = 4.5\% to 233 cm$^{-1}$ at $\epsilon$ = 6\%. The phonon frequencies become negative for $\epsilon$ $>$ 5\% and are pronounced at M point of the Broullion zone. \\

\begin{figure}
\centering
\renewcommand{\thefigure}{S\arabic{figure}}
\includegraphics[width=9cm,height=12cm,keepaspectratio]{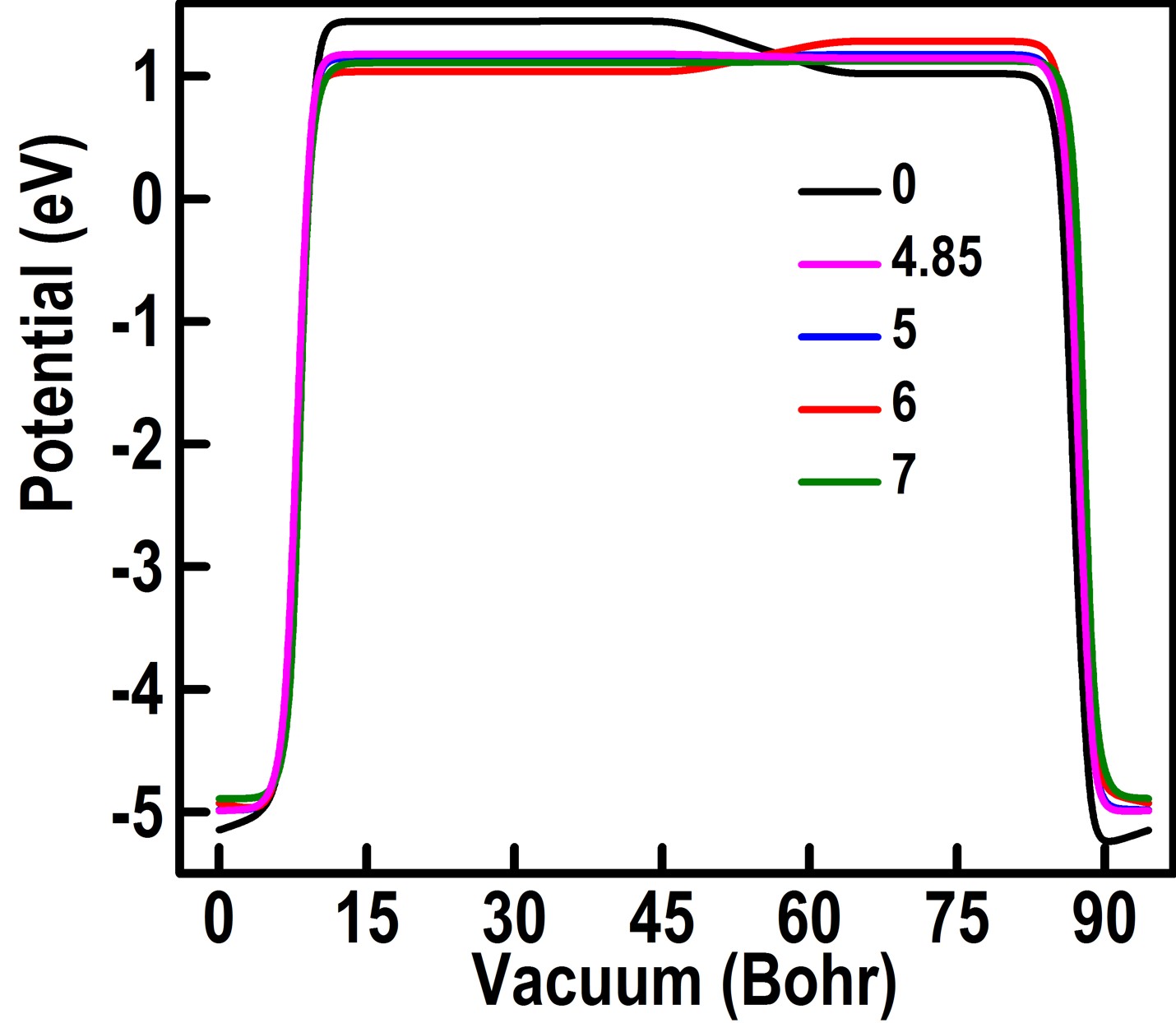}
\caption{Macroscopic average potential of h-NbN as a function of $\epsilon$. }
\end{figure}

Fig. S4 shows the potential of h-NbN w.r.t $\epsilon$. The potential for the unstrained h-NbN between the Nb and N atom differ by 0.42 eV. With the application of strain, the potential on the N atom (V $^N$) decreases and while the potential on the (V $^{Nb}$) increases. At $\epsilon$ = 4.85\%, the difference in the potential between the two atoms is 0.03 eV. While from $\epsilon$ = 7\% - 8\%, the difference is less than 0.013 eV. Such low values are obtained because of the planar configuration of h-NbN from $\epsilon$ = 7\% - 8\%.

\begin{figure}
\centering
\renewcommand{\thefigure}{S\arabic{figure}}
\includegraphics[width=15cm,height=24cm,keepaspectratio]{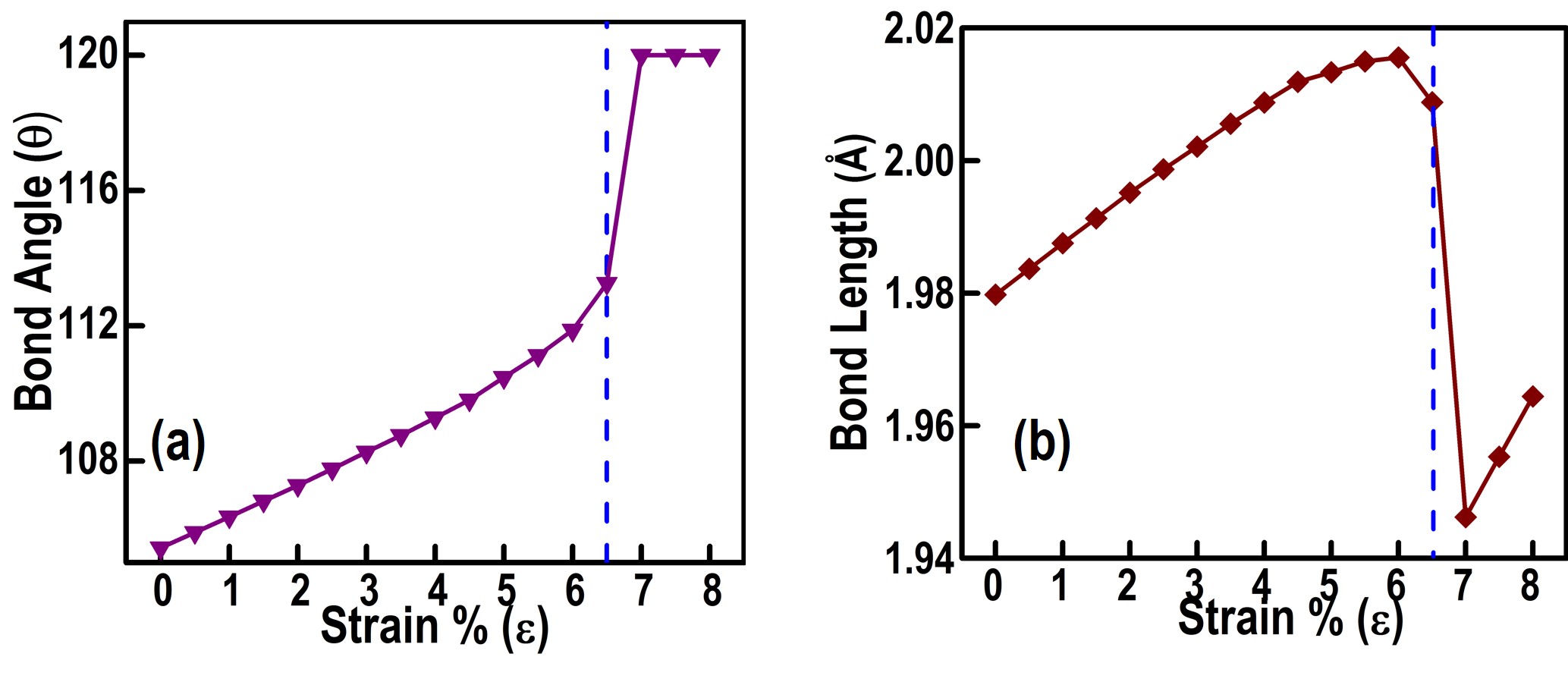}
\caption{(a) Bond angle and (b) bond length of h-NbN as a function of $\epsilon$. The blue dashed line at $\epsilon$ = 6.5\% differentiates the buckled and planar region. }
\end{figure}

Fig. S5 (a) and (b) show the bond angle and bond length  of h-NbN as a function of $\epsilon$ respectively. The bond angle increases to  120$^\circ$ beyond $\epsilon$ = 6.5\% transforming buckled h-NbN to the planar configuration. Unstrained h-NbN possesses an sp$^2$-sd$^2$ hybridization with a bond angle of 105.4$^\circ$  which originates from a compromise between the bonding preferences of Nb (sd$^2$ = 90$^\circ$) and N (sp$^2$ = 120$^\circ$) atoms. This angle is smaller than the angle obtained from sp$^3$ (109.5$^\circ$) hybridized orbitals. With an increase in $\epsilon$, the N atom based sp$^2$ hybridization dominates over the Nb atom based hybridization leading to an increase in the angle. The transition in the structural parameters (bond length and bond angle) is observed primarily beyond $\epsilon$ = 6.5\%. \\

\begin{figure}
\centering
\renewcommand{\thefigure}{S\arabic{figure}}
\includegraphics[width=9cm,height=12cm,keepaspectratio]{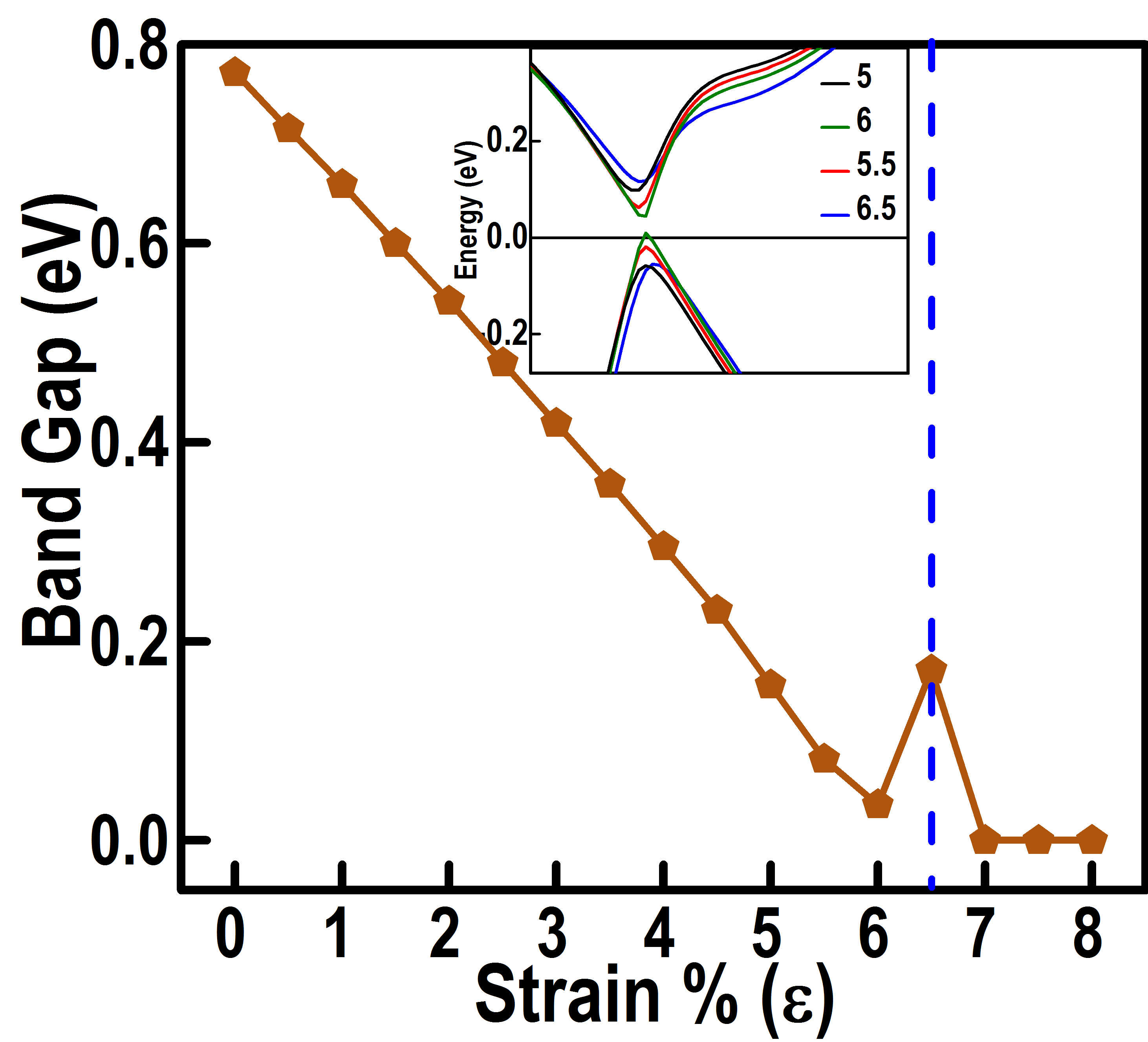}
\caption{Band gap of h-NbN as a function of $\epsilon$. The inset shows the electronic structure with $\epsilon$ varying from 5\% - 6.5\%.}
\end{figure}

We study the electronic structure of h-NbN with respect to $\epsilon$ in Fig. S6. From the bandgap versus $\epsilon$ curve, we see that the bandgap decreases as $\epsilon$ increases from 0 - 6\% and  there is a sudden increase in the value of the bandgap at $\epsilon$ = 6.5\%. For $\epsilon$ = 7\% - 8\%, bandgap of h-NbN reduces to zero. The inset in Fig. S6 shows the bandstructure at $\epsilon$ = 6.5\% where a transition is observed. It can be clearly seen that at $\epsilon$ = 6.5\% the dispersion curve moves apart generating a higher value of bandgap.\\

\begin{figure}
\centering
\renewcommand{\thefigure}{S\arabic{figure}}
\includegraphics[width=15cm,height=24cm,keepaspectratio]{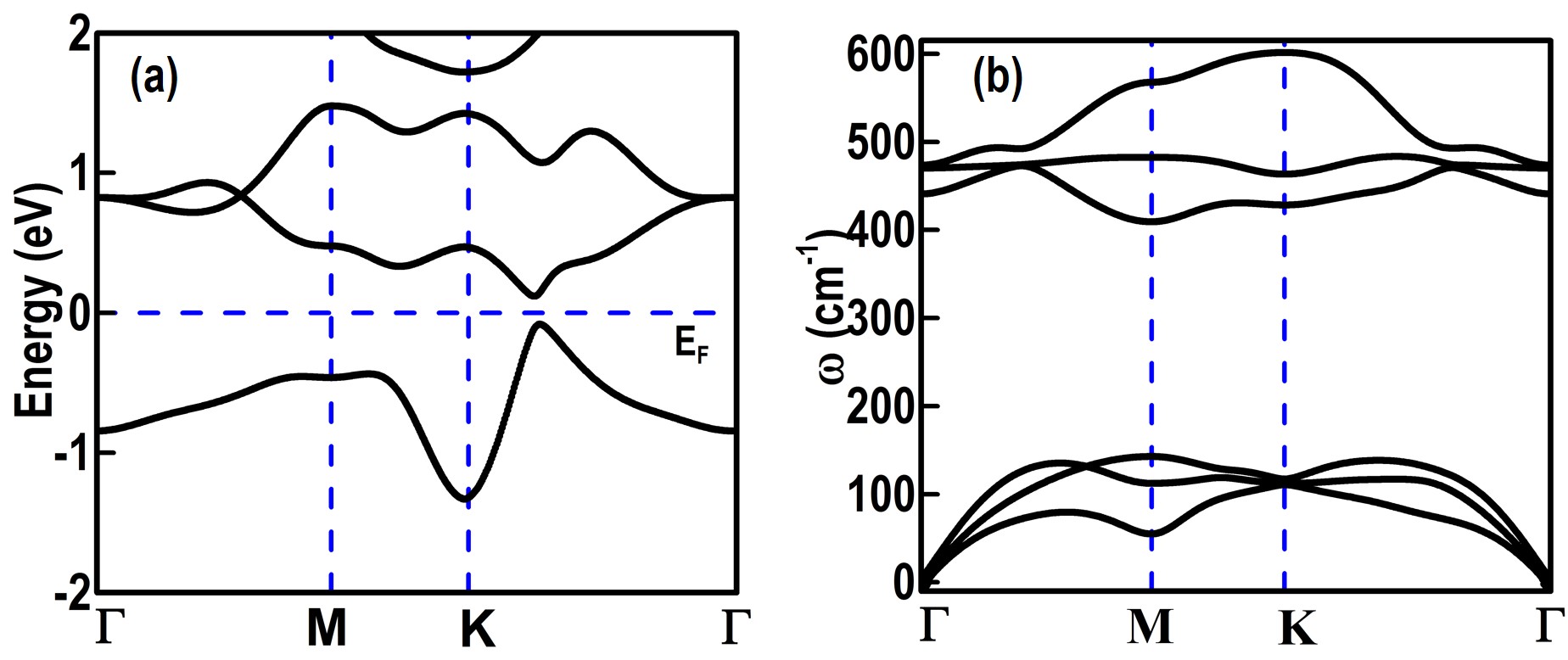}
\caption{(a) Electronic structure and (b) phonon dispersion of h-NbN at $\epsilon$ = 4.7\%. }
\end{figure}

Fig. S7 shows the band structure and phonon dispersion of h-NbN at $\epsilon$ = 4.7\%. It has a direct band gap of 0.2 eV along K $\longrightarrow$ $\Gamma$. The phonon dispersion shows no unstable modes, establishing that the structure is dynamically stable. Branches of its optical and acoustic modes are separated by 267 cm$^{-1}$. \\

 \begin{table}
        \renewcommand{\thetable}{S\arabic{table}}
     
  \begin{center}     
    \caption{Comparison of energetics and buckling of metastable and transition states (Fig. 2(c)) seen in the triple well energy landscape of h-NbN at $\epsilon$ = 4.7\% and $\epsilon$ = 5\%. }
    \label{tab:table1}
    \begin{tabular}{l|c|r} 
      \textbf{Strain ($\epsilon$)} & \textbf{4.7\%} & \textbf{5\%}\\
      \hline
      E$_2$-E$_3$ (eV) & 0.045 & 0.057\\
      E$_3$-E$_1$ (eV) & 0.078 & 0.035\\
      E$_2$-E$_1$ (eV) & 0.092 & 0.123\\
      $\Delta$$_1$ (\AA) & -0.65 & -0.637\\
      $\Delta$$_2$ (\AA)\let\thefootnote\relax\footnote {The values of $\Delta$$_4$ and $\Delta$$_5$ are opposite in sign w.r.t $\Delta$$_2$ and $\Delta$$_1$ respectively.} & -0.318 & -0.327 \\
    \end{tabular}
  \end{center}
\end{table}

\begin{figure}
\centering
\renewcommand{\thefigure}{S\arabic{figure}}
\includegraphics[width=15cm,height=24cm,keepaspectratio]{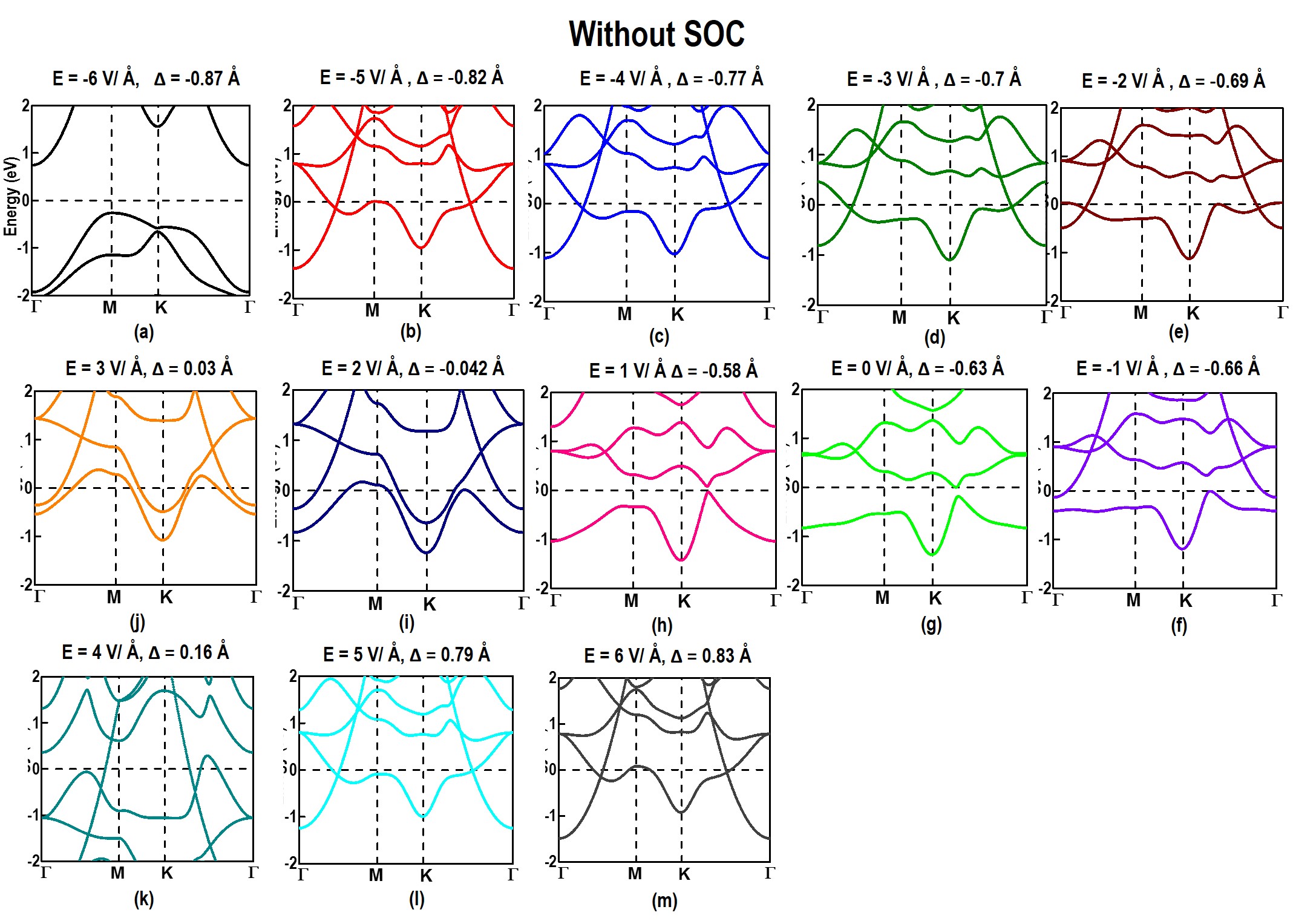}
\caption{(a) - (m) Electronic structure of h-NbN at $\epsilon$ = 5\% under the effect of E (-6 V/\AA \hspace{0.07 cm} to 6 V/\AA \hspace{0.07 cm} forward path (red curve)). The value of electric field and corresponding buckling is specified on the top of each panel.}
\end{figure}

Fig. S8 and S9 show the electronic structure of h-NbN at $\epsilon$ = 5\% with the application of electric field along the forward path (red curve in Fig. 3(a)) without and with the effect of spin-orbit coupling (SOC). The electronic structure of h-NbN exhibits both Dirac and gapped semi-metallic states with respect to the electric field. We note that Dirac semi-metallic state occurs at relatively large electric fields. One of the bands crossing Dirac point is constituted of d-orbitals of Nb, while the other one is a parabolic free-electron like band, which arises from the displacement of small electronic charge from h-NbN to vacuum. While the former band is spin-split, the latter remains spin-degenerate, and this results in triple point semi-metallic states. \\

\begin{figure}
\centering
\renewcommand{\thefigure}{S\arabic{figure}}
\includegraphics[width=15cm,height=24cm,keepaspectratio]{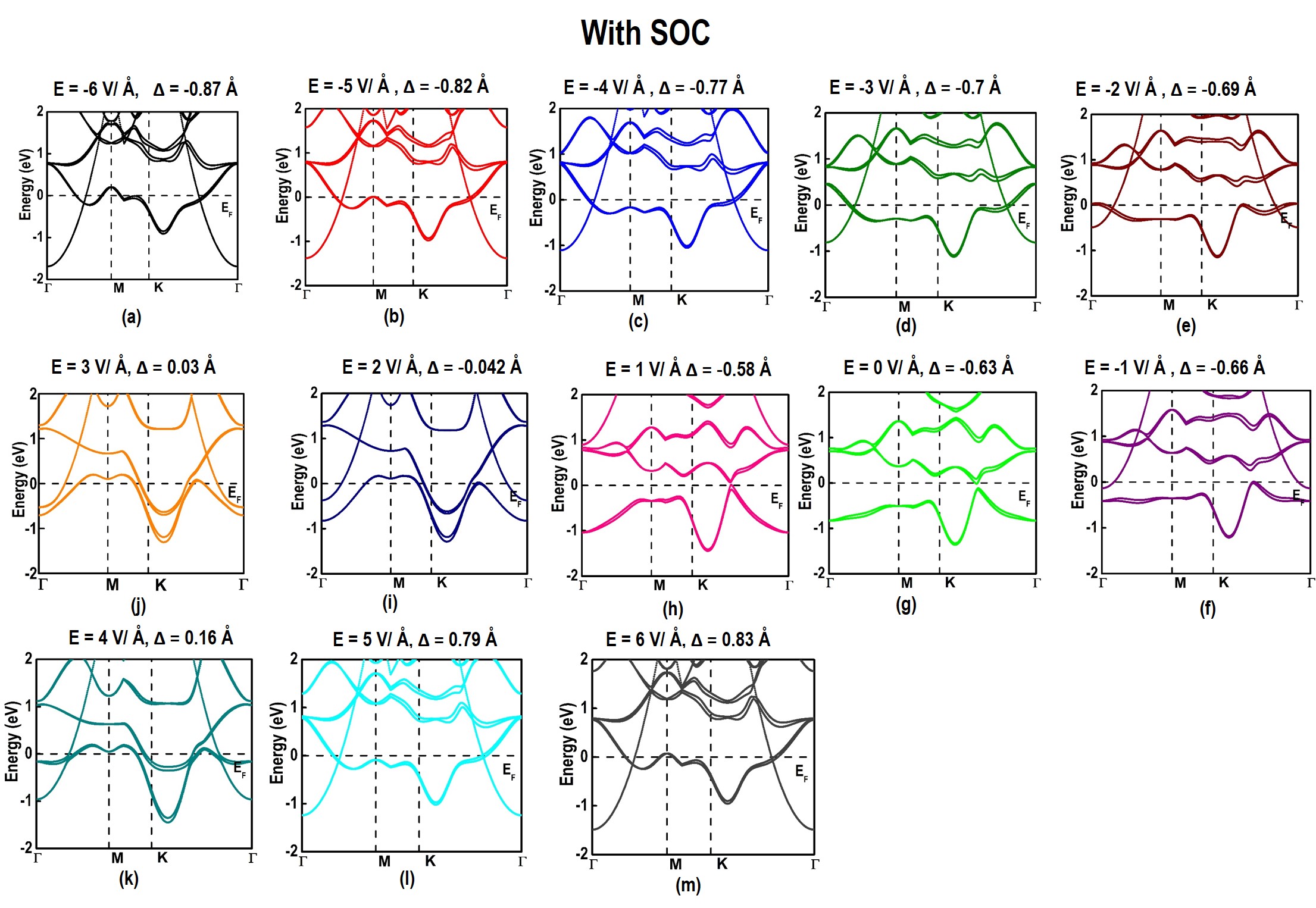}
\caption{(a) - (m) Electronic structure of h-NbN at $\epsilon$ = 5\% under the effect of E (-6 V/\AA \hspace{0.07 cm} to 6 V/\AA \hspace{0.07 cm} forward path (red curve)) with the effect of SOC. The value of electric field and corresponding buckling is specified on the top of each panel.}
\end{figure}

\begin{figure}
\centering
\renewcommand{\thefigure}{S\arabic{figure}}
\includegraphics[width=15cm,height=24cm,keepaspectratio]{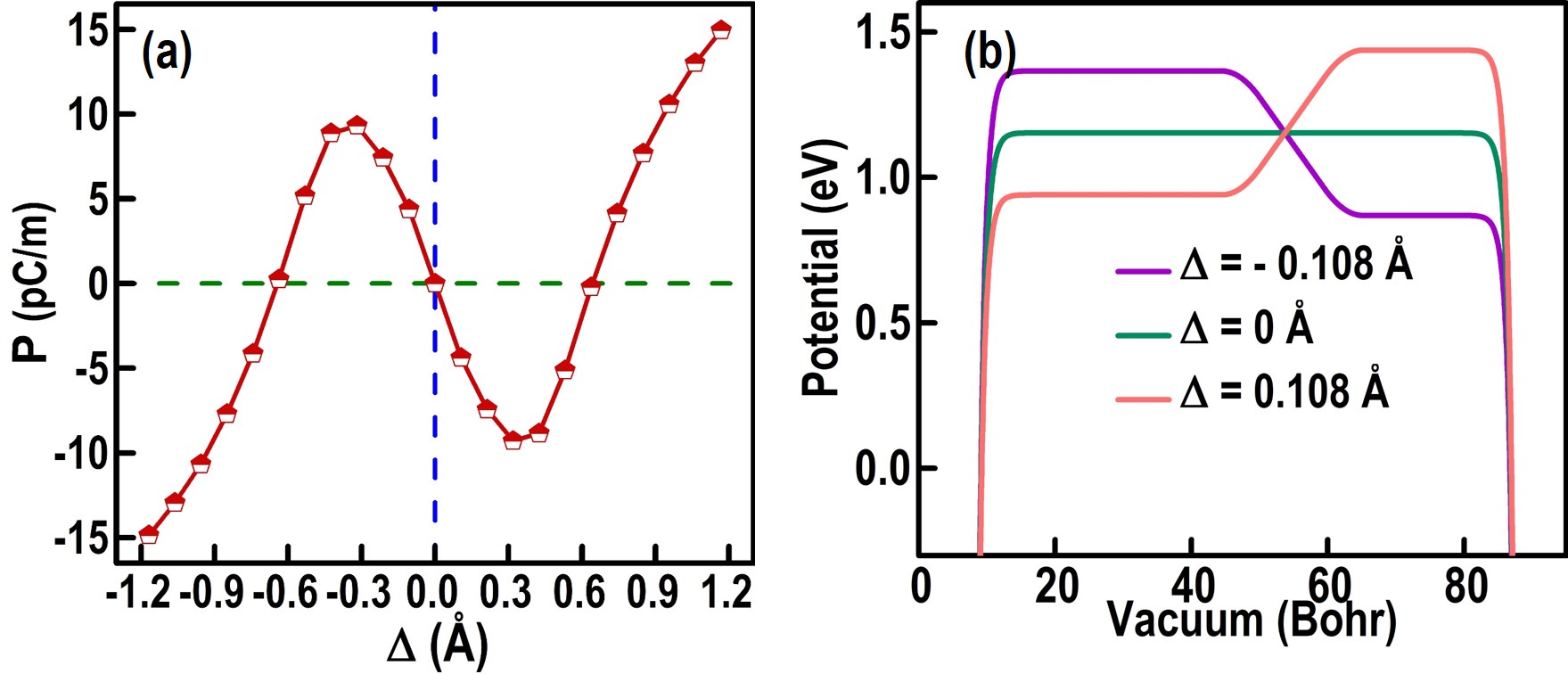}
\caption{(a) Polarization versus buckling at E = 0 V/\AA \hspace{0.07 cm} for h-NbN at $\epsilon$ = 5\%, (b) macroscopic average potential at E = 0 V/\AA \hspace{0.07 cm}. }
\end{figure}

Fig. S10 (a) shows polarization as a function of buckling of h-NbN (corresponding to triple well potential in Fig. 2 (c)) for $\epsilon$ = 5\% at E = 0 \hspace{0.07 cm}V/\AA. We notice that polarization changes sign with buckling and becomes zero with V(z) becoming flat (Fig. S10 (b)) for $\Delta$ = 0\hspace{0.07 cm}\AA.

\begin{figure}
\centering
\renewcommand{\thefigure}{S\arabic{figure}}
\includegraphics[width=15cm,height=24cm,keepaspectratio]{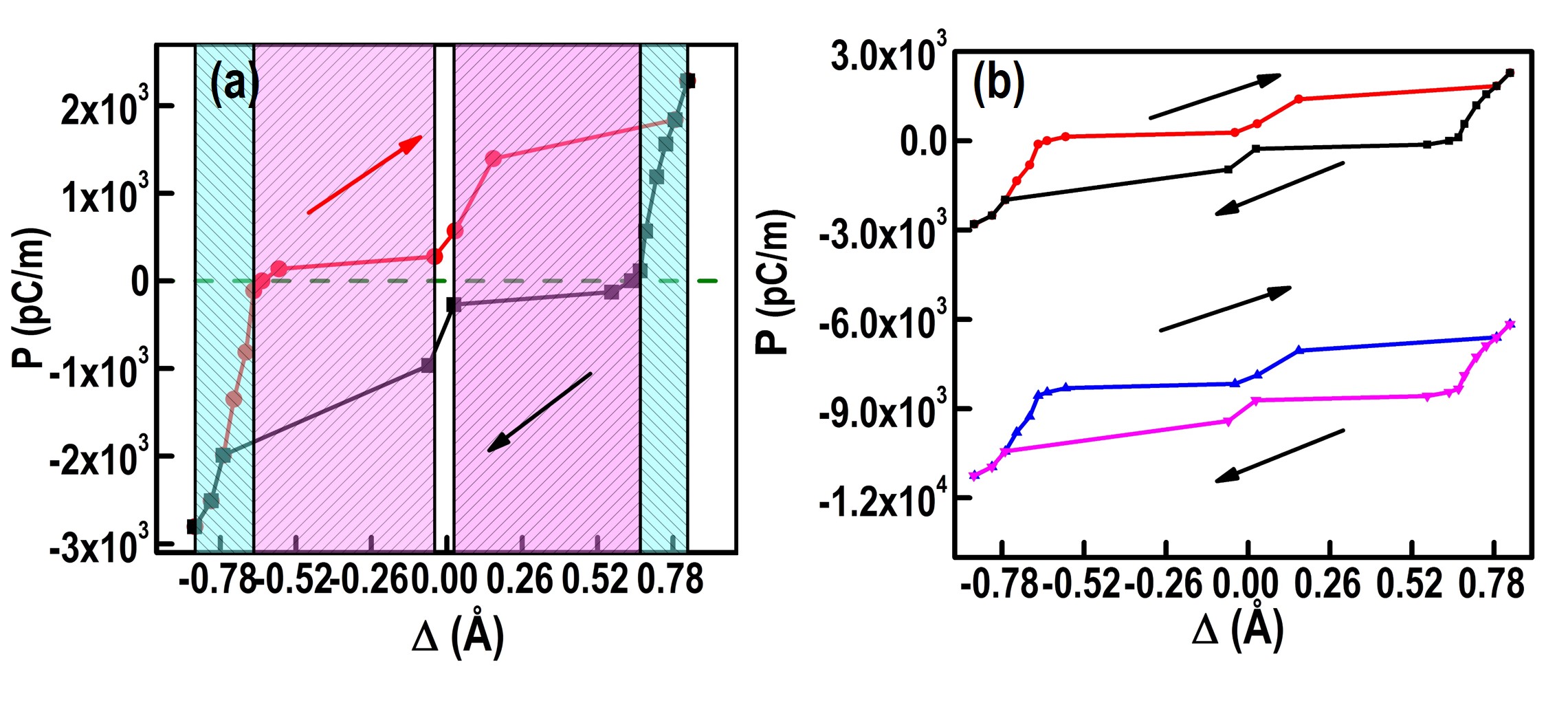}
\caption{(a) P($\Delta$, E) curve with polarization quanta and (b) branches of P($\Delta$, E) curve for h-NbN at $\epsilon$ = 5\%.}
\end{figure}

We show P($\Delta$, E) curve with polarization quanta in Fig. S11 (a) and its branches in Fig. S11 (b). We observe an asymmetry in the ionic and electronic responses of h-NbN. For $\Delta$ $<$ 0 \hspace{0.07 cm}\AA, a negative electric field results in a significant change in polarization without notable change in $\Delta$ (see the blue shaded region in Fig. S11 (a)), while a positive electric field changes the structure ($\Delta$) remarkably without changing the polarization significantly (see the pink shaded region in Fig. S11 (a)). Clearly, electrons screen changes in $\Delta$ and yield a small polarization when E $>$ 0, while they give a large polarization themselves in response to electric field, when $\Delta$ does not change much for E $<$ 0.  

\begin{figure}
\centering
\renewcommand{\thefigure}{S\arabic{figure}}
\includegraphics[width=15cm,height=24cm,keepaspectratio]{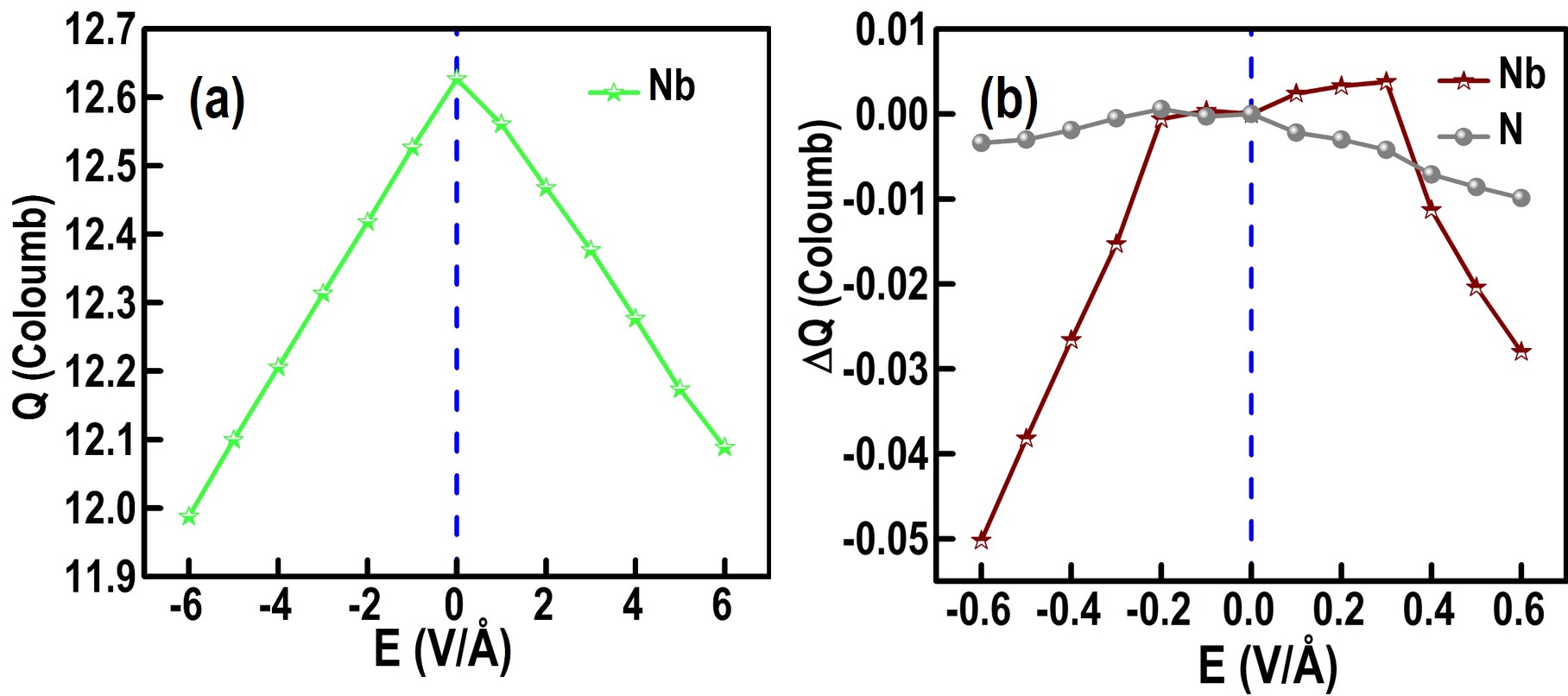}
\caption{(a) Lowdin charges on the Nb atom for the electric field in the range -6 V/\AA \hspace{0.07 cm} to 6 V/\AA \hspace{0.07 cm} (b) change in the Lowdin charges for electric field in the range -0.6 V/\AA \hspace{0.07 cm} to 0.6 V/\AA \hspace{0.07 cm}, of h-NbN at $\epsilon$ = 5\%.  }
\end{figure}

Fig. S12 (a) and (b) show the Lowdin charges on the Nb atom for high electric field and changes in Lowdin charges on both Nb and N atom for low electric field values. We observe an asymmetry in Lowdin charges on Nb atom both for high and low values of electric field. The change in the Lowdin charges at low field values is of opposite sign, which indicates that electronic ferroelectrcity is caused by the charge transfer. However, we point out that the estimation of charges in pd-hybridized systems is tricky and these subtle changes in charges can have nontrivial consequences.

\begin{figure}
\centering
\renewcommand{\thefigure}{S\arabic{figure}}
\includegraphics[width=15cm,height=24cm,keepaspectratio]{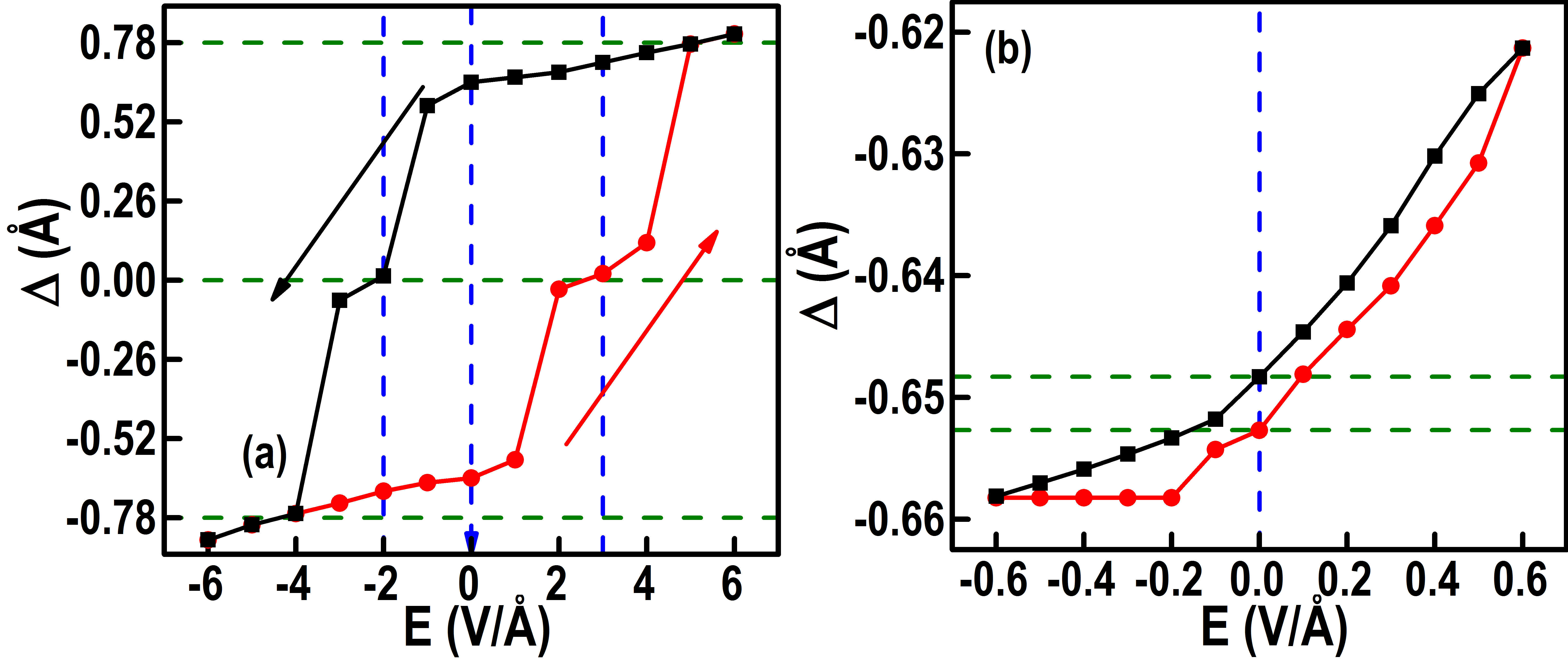}
\caption{Buckling versus electric field in the range (a) -6 V/\AA \hspace{0.07 cm} to 6 V/\AA \hspace{0.07 cm} (b) -0.6 V/\AA \hspace{0.07 cm} to 0.6 V/\AA \hspace{0.07 cm} of h-NbN at $\epsilon$ = 4.7\%.}
\end{figure}

\begin{figure}
\centering
\renewcommand{\thefigure}{S\arabic{figure}}
\includegraphics[width=9cm,height=12cm,keepaspectratio]{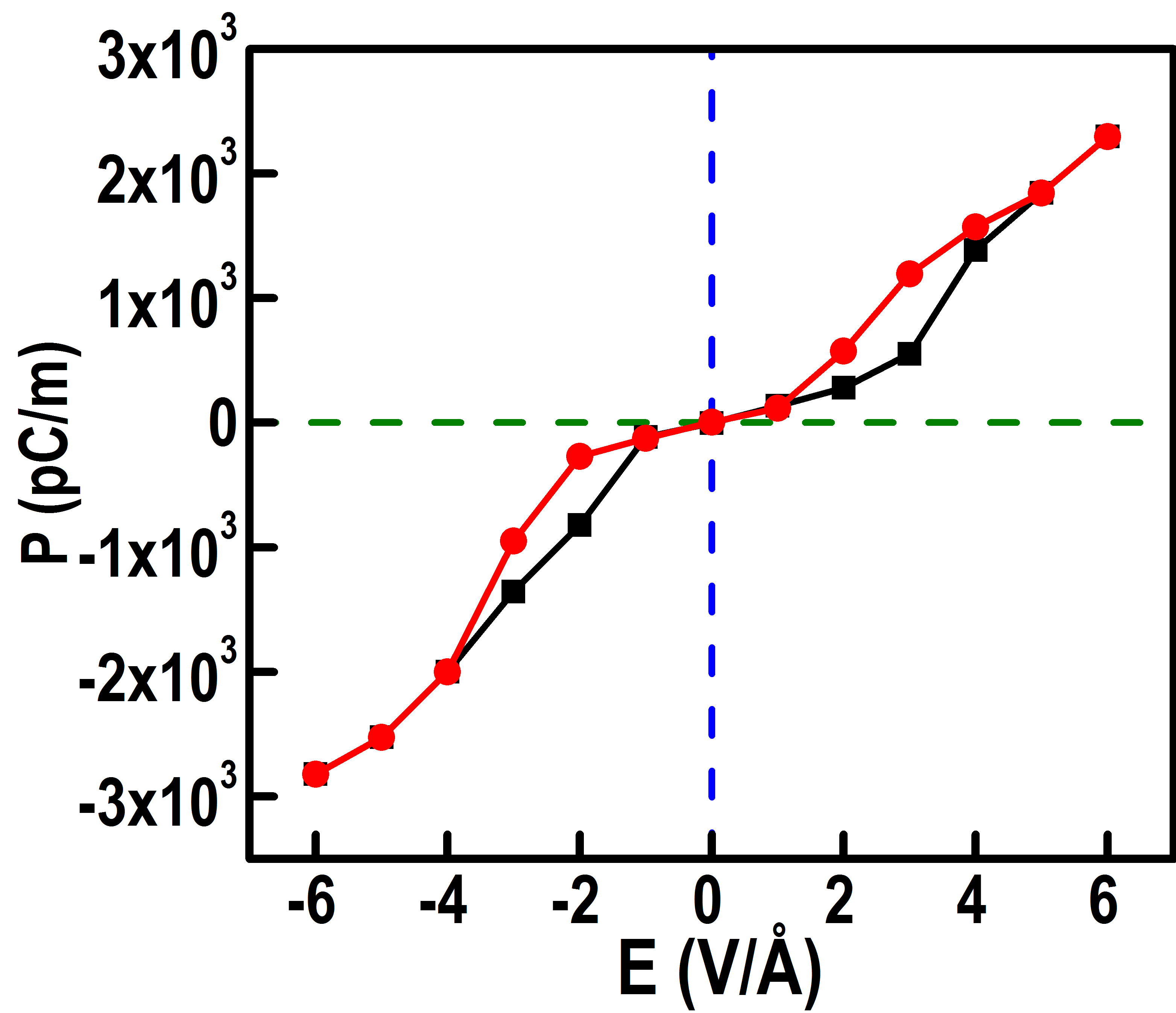}
\caption{Polarization as a function of electric field  of h-NbN at $\epsilon$ = 4.7 \%.}
\end{figure}

The buckling and polarization of h-NbN at $\epsilon$ = 4.7\% are shown in Fig. S13 and S14. The hysteresis (both in buckling and polarization) in h-NbN at $\epsilon$ = 4.7\% is similar to $\epsilon$ = 5\% (see Fig. 3(a) and Fig. 3(b)).

\begin{figure}
\centering
\renewcommand{\thefigure}{S\arabic{figure}}
\includegraphics[width=9cm,height=12cm,keepaspectratio]{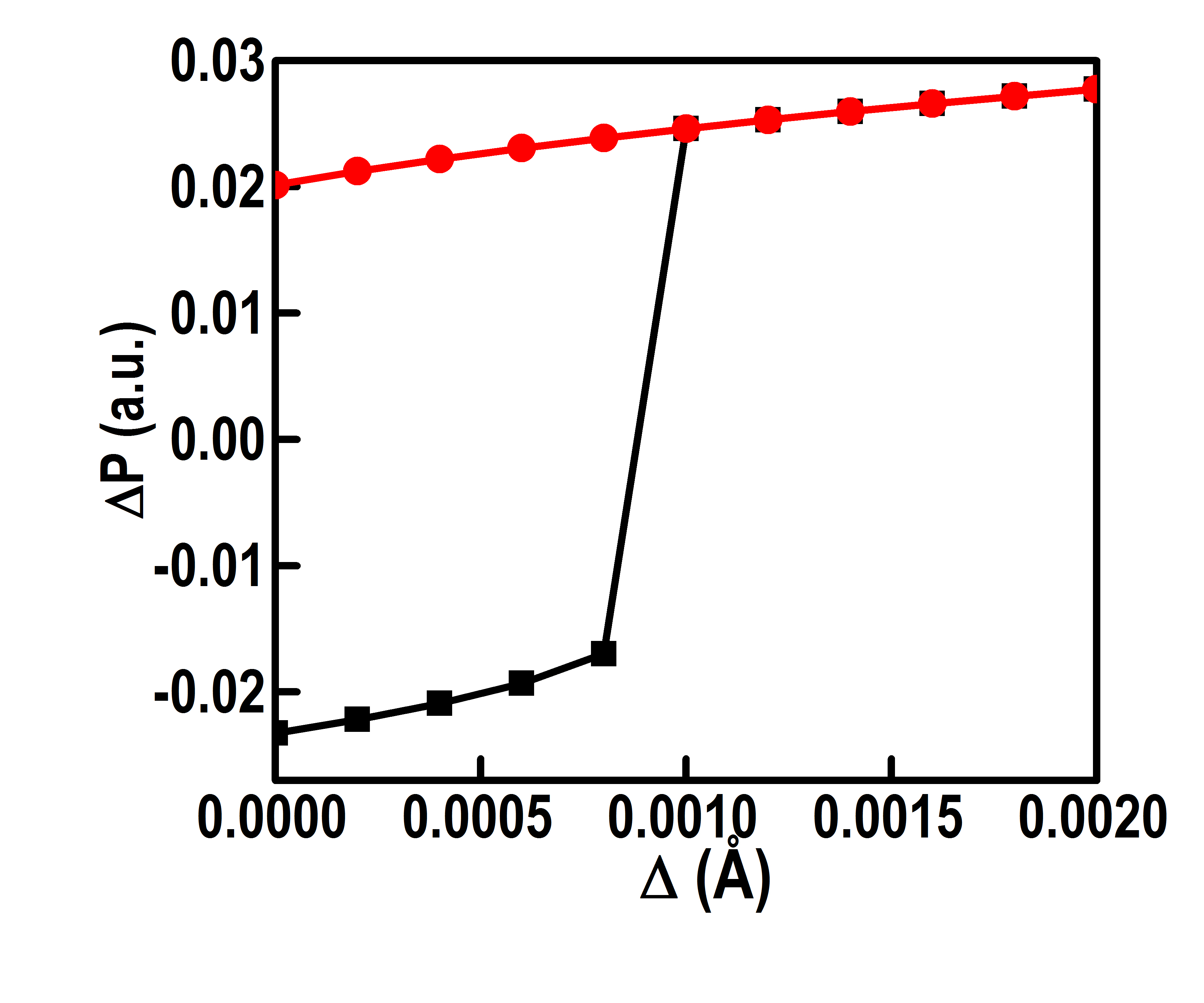}
\caption{(a) Polarization as a function of buckling obtained using our model Hamiltonian with the cubic terms in eq.1 in the manuscript.}
\end{figure}